\documentclass[%
 reprint,
 amsmath,amssymb,
 prl,
]{revtex4-2}
\usepackage{graphicx}
\usepackage{bm}
\usepackage{hyperref}
\usepackage{amsmath}
\begin{document}
\title{Universal Statistics of Energy and Information Flow\\in Random Electromagnetic Fields}
\author{Yuchen Ke$^{1}$}
\author{Nandini Bhattacharya$^{1}$}
\author{Stefan Rotter$^{2}$}
\author{Fabian Maucher$^{1}$}
\email{f.maucher@tudelft.nl}
\affiliation{
$^1$Department of Precision and Microsystems Engineering, Faculty of Mechanical Engineering, Delft University of Technology, 2628 CD Delft, The Netherlands\\
$^2$Institute for Theoretical Physics, Vienna University of Technology, Vienna, A-1040, Austria
}
\begin{abstract}
We establish a universal statistical description for the local flow of energy and information in random electromagnetic fields. The longitudinal Poynting flux, written as a Hermitian quadratic form of the transverse electric and magnetic field components, follows a probability distribution that is completely determined by four eigenvalues of an electromagnetic covariance matrix. These flux eigenvalues quantify forward transport, optical backflow, and polarization mixing, and reduce to the known paraxial and isotropic limits in the appropriate regimes – including strongly nonparaxial fields, where no universal description was known so far. Full-vector simulations of continuous and discrete disordered media confirm this universality. The same framework applies to the recently introduced Fisher-information flux, with the fields replaced by their sensitivity to a parameter, thereby unifying the statistics of local energy and information transport in random light and revealing the reversal of information flow across a parameter-dependent object.
\end{abstract}
\maketitle
Speckle is one of the most familiar manifestations of wave interference in disordered media~\cite{goodman2007speckle,Goodman2015}, and its most remarkable feature is universality: once a scalar field has been fully randomized by multiple scattering, its complex amplitude is Gaussian distributed and the intensity follows the Rayleigh law, independent of the microscopic realization of the disorder~\cite{Goodman2015}. Departures from this law can be engineered by introducing correlations~\cite{Cao:PRL:2014,PhysRevLett.127.093903, han2023tailoring}, but the Gaussian regime remains the reference point of statistical optics, underpinning applications from wavefront shaping and imaging through opaque media to compact spectrometry and precision displacement sensing~\cite{Vellekoop2007,Vellekoop2008,Popoff:prl:2010,Redding2013,vanPutten2012,Rotter2017,Bouchet2021}.

Intensity, however, only quantifies the local energy density of the light field; it does not describe how energy flows. The transport of electromagnetic energy is encoded in the Poynting vector, and for general nonparaxial fields – as generated by strong scattering, near fields, or high-numerical-aperture optics – this flux decouples from the intensity: all vector components contribute, the electric and magnetic fields are not locked to a single propagation direction, and the local Poynting vector can display vortices and negative longitudinal flux even when the net energy transport remains forward~\cite{berry2010quantum,kotlyar2018energy,yuan2019plasmonics}. 

Whether the universality of speckle intensity statistics extends to local electromagnetic energy currents has remained an open question. The answer is known only in two limiting cases. In paraxial speckle, the longitudinal Poynting flux is proportional to the intensity and hence strictly positive, following Rayleigh or gamma statistics~\cite{Goodman2015}. In the opposite limit of an isotropic monochromatic random vector field, the average flux vanishes and the momentum distribution becomes symmetric between forward and backward directions~\cite{gadeyne2024optical}. Related statistics of quadratic field quantities have also been studied for isotropic reverberant fields in other areas of wave physics, e.g., for the active and reactive sound intensity in room acoustics~\cite{Ebeling1984,Jacobsen2011} and for the energy density and received power in electromagnetic reverberation chambers~\cite{Hill2009}; all of these treatments, however, rest on statistical isotropy. However, most random optical fields encountered in scattering experiments are neither paraxial nor isotropic: they retain a finite directed energy current while also containing local backflow and polarization-dependent electromagnetic correlations.
\begin{figure}[t]
\includegraphics[width=0.48\textwidth]{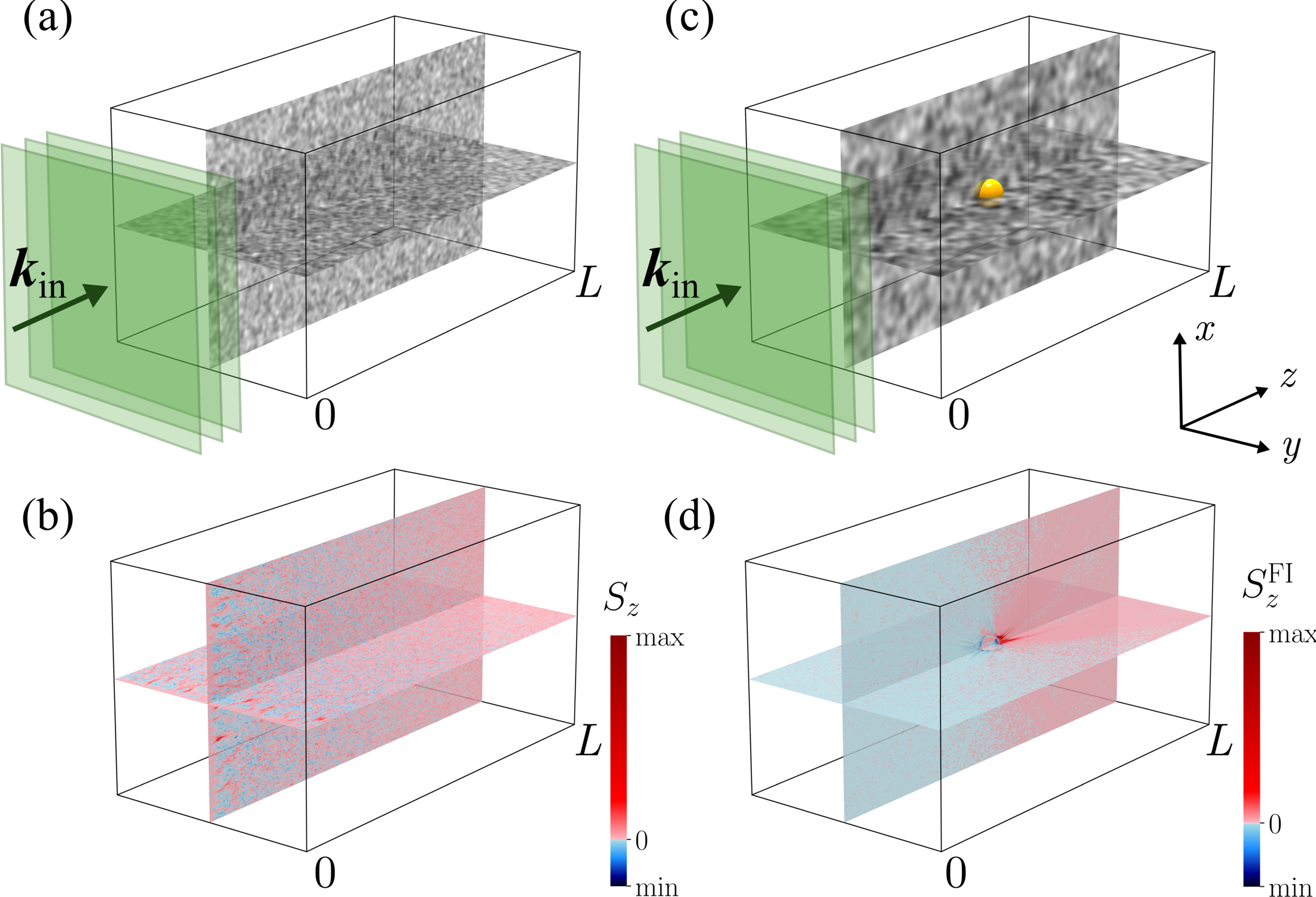}
\caption{
Concept of local flux statistics in vectorial speckle.
(a) A coherent plane wave (green wavefronts) is launched from a homogeneous region into a finite disordered slab extending from $z=0$ to $z=L$. The gray textured planes denote representative disorder cross sections.
(b) Inside the slab, multiple scattering converts the initially ballistic field into vectorial speckle with local domains of positive (red) and negative (blue) longitudinal Poynting flux $S_z$.
This applies more broadly: Illumination of spherical scatterer (yellow) embedded in the disordered medium in (c) acts as a source of the Fisher-information flux (d).
\label{fig:concept}}
\end{figure}

Here we show that the Gaussian random-field regime of vectorial multiple scattering is governed by a compact universal law: the full probability distribution of the longitudinal Poynting flux is fixed by four eigenvalues of an electromagnetic covariance matrix. These flux eigenvalues provide a local counterpart of transport eigenvalues in mesoscopic scattering~\cite{Dorokhov1982JETPL,MELLO1988,Beenakker1997RMT,Rotter2017,g3kd-sg4x,djhy-16mh}: instead of characterizing channel-resolved transmission, they quantify the statistical weights of forward flow, backflow, and polarization mixing in vectorial speckle.

The same idea naturally extends from energy flow to information flow.
The recently introduced Fisher-information flux describes the spatial transport of information about a parameter of a scattering system~\cite{Hpfl2024,weimar2026controllingflowinformationoptical}.
As we show below, this information current obeys the same eigenvalue statistics as the Poynting flux, with the physical fields replaced by their  sensitivity to the parameter.
This provides a unified statistical description of local energy and information transport in random electromagnetic fields.

As a starting point, we consider a monochromatic field in a linear, nonmagnetic medium,
\begin{equation}
    \nabla \times \nabla\times{\bm  E}({\bm  r}) - k({\bm  r})^2 {\bm  E}({\bm  r}) = {\bm 0} 
    \label{eq:vector_wave_eq}
\end{equation}
with local wavenumber $k({\bm r})=2\pi n({\bm r})/\lambda$.
The disordered region is a finite slab in the range $0<z<L$, where the refractive index is
$n({\bm r})=n_0+\eta({\bm r})$, with $\eta({\bm r})$ denoting spatially varying random fluctuations.
Outside this interval, the medium is homogeneous with refractive index $n_0$.
A monochromatic $x$-polarized plane wave is launched from the homogeneous input region $z<0$ and propagates into the disordered slab, where repeated multiple scattering progressively converts the initially ballistic wave into a vectorial speckle field [cf. Fig.~\ref{fig:concept}(a,b)]. 
The flux statistics discussed below are evaluated at transverse planes inside or beyond the disordered region once multiple scattering has generated a Gaussian random-field regime.
This regime is the vectorial analog of the fully developed scalar speckle considered in classical speckle theory~\cite{Goodman2015}, but it does not require the field to be statistically isotropic.
In particular, the average longitudinal Poynting flux may remain finite, reflecting the preferred propagation direction inherited from the incident beam.

The central statistical assumption of our theory concerns the transverse field vector
\begin{equation}
{\bm Z}=(E_x,E_y,H_x,H_y)^{\rm T}\,,
\end{equation}
whose four components fully determine the longitudinal Poynting flux $S_z=\frac{1}{2}\operatorname{Re}\{E_xH_y^*-E_yH_x^*\}$. We assume that $\bm Z$ is a multivariate complex Gaussian random variable; this is expected to hold once each field component results from the coherent superposition of a large number of statistically independent scattering paths with effectively random phases. According to the multivariate central limit theorem, the field components then approach a joint Gaussian distribution~\cite{Goodman2015}.
Importantly, this does not require the components to be independent.
All electromagnetic correlations between electric and magnetic fields are retained through the covariance matrix
\begin{equation}
{\bm C}=\langle{\bm Z}{\bm Z}^{\dagger}\rangle .
\end{equation}
Two further properties of fully developed speckle make $\bm C$ the complete statistical description. First, $\bm Z$ has zero mean, $\langle {\bm Z}\rangle={\bm 0}$, since the coherent (ballistic) component of the field decays within a few scattering mean free paths $\ell_s$. Second, $\bm Z$ is circular, $\langle{\bm Z}{\bm Z}^{\rm T}\rangle={\bm 0}$: because each field component is a superposition of many partial waves with uniformly distributed phases, the ensemble is invariant under a global phase shift ${\bm Z}\rightarrow e^{i\varphi}{\bm Z}$, which enforces both properties simultaneously. The only statistical assumption entering the theory is therefore zero-mean circular Gaussianity of the vector ${\bm Z}$; the covariance matrix itself is arbitrary and contains the nonparaxial and polarization-dependent structure of the field.

The longitudinal Poynting flux $S_z=\frac{1}{2}\operatorname{Re}\{E_xH_y^*-E_yH_x^*\}$ we are interested in, can now be written as
\begin{equation}
S_z  ={\bm Z}^\dagger{\bm A}{\bm Z},
\label{eq:quadratic_form}
\end{equation}
where $\bm A$ is an anti-diagonal matrix with elements $(1,-1,-1,1)/4$.
Since the longitudinal Poynting flux is a Hermitian quadratic form of a zero-mean circular complex Gaussian field, its characteristic function follows directly from the classical result of Turin for quadratic forms of complex Gaussian random variables~\cite{turin1960characteristic}. 
We thus obtain the probability density
\begin{equation}
\resizebox{0.98\columnwidth}{!}{$\displaystyle
P(S_z)=
\int_{-\infty}^{\infty}
\frac{\mathrm{e}^{-\mathrm{i}kS_z}\,dk}
{2\pi\det(\mathbb{I}-\mathrm{i}k{\bm C}{\bm A})}
=
\int_{-\infty}^{\infty}
\frac{\mathrm{e}^{-\mathrm{i}kS_z}\,dk}
{2\pi\prod_{n=1}^{4}(1-\mathrm{i}k\mu_n)}
$}.
\label{eq:statistics}
\end{equation}
Here, $\mu_n$ are the four real eigenvalues of ${\bm C}{\bm A}$.
Because $\operatorname{Tr}({\bm C}{\bm A})=\langle S_z\rangle$, their sum gives the mean longitudinal energy flux. Equation~(\ref{eq:statistics}) is the central result of this Letter: for any zero-mean circular vectorial Gaussian random field, all details of the local Poynting-flux distribution are completely determined by the four flux eigenvalues $\mu_n$.
For nondegenerate eigenvalues, Eq.~(\ref{eq:statistics}) yields the closed form
\begin{equation}
P(S_z)=\sum_{n=1}^4 \alpha_n\mathrm{e}^{-\frac{S_z}{\mu_n}}\Theta(\mu_nS_z), \quad
\alpha_n
= \frac{\mu_n \lvert \mu_n \rvert}
       {\displaystyle \prod_{\substack{m=1 \\ m \ne n}}^{4} (\mu_n - \mu_m)}\,,
\label{eq:PSz}
\end{equation}
where the $\alpha_n$ are prefactors that can be found via partial fraction decomposition.
Positive eigenvalues $\mu_n >0$ determine the exponential tails for forward energy flow, while negative eigenvalues $\mu_n <0$ determine the tails associated with optical backflow.
The degeneracies and signs of $\mu_n$ identify the physical regime of the field.
For scalar paraxial speckle, the eigenvalues of ${\bm C}{\bm A}$ are
$(\langle S_z\rangle,0,0,0)$ and Eq.~(\ref{eq:PSz}) reduces to the Rayleigh intensity law.
For paraxial unpolarized speckle, they become $(\langle S_z\rangle/2,\langle S_z\rangle/2,0,0)$ and one obtains the usual gamma distribution (see End Matter).
For an isotropic vectorial random field, $\langle S_z\rangle=0$ and the four flux eigenvalues are doubly degenerate and given by $(\mu,\mu,-\mu,-\mu)$ with $\mu=\sqrt{\langle S_z^2\rangle}/2$, yielding
\[
P_{\rm iso}(S_z)
=
\frac{1+|S_z|/\mu}{4\mu}
\mathrm{e}^{-|S_z|/\mu},
\]
in agreement with Ref.~\cite{gadeyne2024optical}.
The general nonparaxial regime interpolates between these limits without being described by either of them.
\begin{figure}[t]
\includegraphics[width=0.48\textwidth]{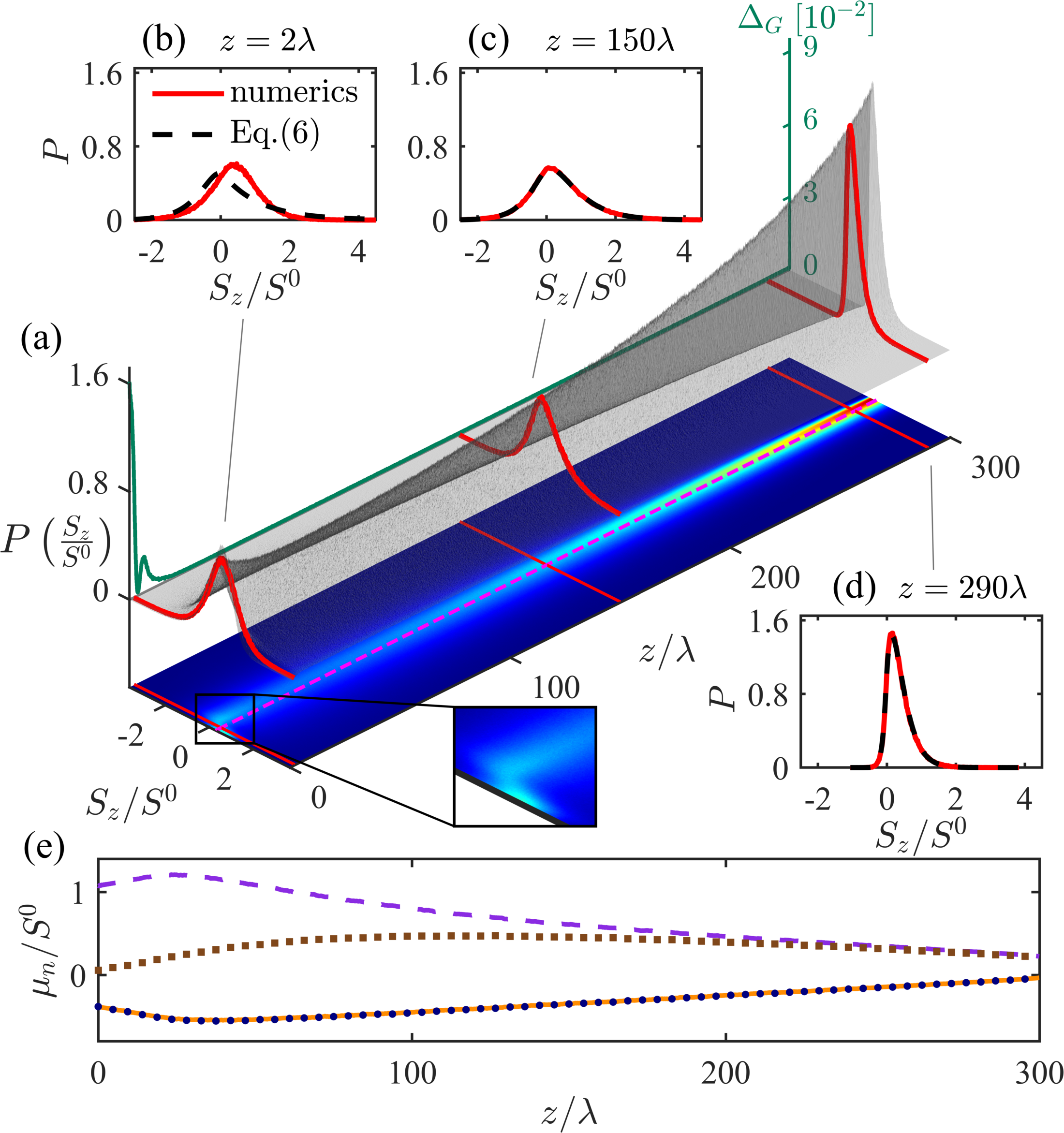}
\caption{
Universal Poynting-flux statistics in a continuous disordered medium.
(a) Evolution of the probability distribution $P(S_z)$ with propagation distance $z$ for an incident linearly polarized plane wave. $S^0$ is the Poynting flux for the homogeneous medium ($\eta=0$) and is used for normalization.
The disordered slab extends from $0$ to $z=300\lambda$; the dashed magenta line in the density plot marks the conserved mean flux $\langle S_z\rangle$.
The inset magnifies the probability distribution over the first few scattering mean free paths, where the electromagnetic field evolves from ballistic propagation to Gaussian speckle.
The second green axis quantifies the deviation  $\Delta_G$ from a Gaussian random-field regime, indicating a gradual onset of the theory's applicability for  $z/\lambda\gtrsim 10$, where $z\gg\ell_s$.
(b--d) Representative slices comparing numerical data with Eq.~(\ref{eq:PSz}); (b) corresponds to $z\simeq 1.5\ell_s$, before the onset of Gaussian statistics and applicability of our theory.
(e) Evolution of the four flux eigenvalues $\mu_n(z)$ of ${\bm C}{\bm A}$.
The positive (negative) eigenvalues determine the forward-flow (back-flow) contributions to the distribution.
\label{fig:poynting}}
\end{figure}

To test Eq.~(\ref{eq:PSz}), we solve Eq.~(\ref{eq:vector_wave_eq}) using a modified Born-series method~\cite{osnabrugge:JCompPhys:2016,osnabrugge:OpEx:2021}.
Absorbing boundaries are used along $z$ and periodic boundaries in the transverse directions mimic bulk propagation.
Ensemble averages are obtained by transverse spatial averaging.
We compute the eigenvalues $\mu_n(z)$ by first filling the covariance matrix $\bm{C}$ by evaluating $\langle Z_iZ_j^*\rangle$ using the numerical fields and after that diagonalizing $\bm{CA}$.
We first study a continuous disordered medium with Gaussian correlation function
\begin{equation}
\langle \eta({\bm r})\eta({\bm r}')\rangle=
\frac{\kappa^2}{(2\pi\sigma^2)^{3/2}}
\exp[-({\bm r}-{\bm r}')^2/(2\sigma^2)] .
\end{equation}
For the data in Fig.~\ref{fig:poynting}, the slab begins at $z_{\rm in}=0$ and ends at $z_{\rm out}=L$, with $L=300\lambda$ chosen as in the numerical domain shown in the figure.
We use $\sigma=\lambda$, $\kappa/\sigma^{3/2}=0.4$, and $n_0=1.5$, corresponding to a scattering mean free path $\ell_s=1.2\lambda$ and a transport mean free path $\ell_t=96\lambda$. The fact that we have $\ell_t\gg \ell_s$ reflects the strong forward-scattering of the correlated disorder chosen here \cite{Heller2021BranchedFlow,brandstotter_branched}.
Figure~\ref{fig:poynting}(a) shows how the distribution of $S_z$ develops from an initially broadened ballistic peak into a vectorial-speckle distribution with appreciable negative-flux weight.
The mean flux is conserved during propagation, as required by stationarity and the transverse periodic boundary conditions, but the shape of the distribution changes strongly.
Once the field is sufficiently Gaussian, Eq.~(\ref{eq:PSz}) accurately captures the full distribution, including both positive and negative tails [Figs.~\ref{fig:poynting}(b--d)].
We quantify the deviation from Gaussianity $\Delta_G$ by evaluating $\Delta_G = \sum_{j}\int \left[P(A_{j})-P_{\mathrm{theo}}(A_{j})\right]^2 dA_j$, where $j=x,y,z$ denotes the component,  $A_j$ is a normalized field amplitude $A_j = |E_j|/\sigma_{|E_j|}$ and the theoretical distribution is given by $P_{\mathrm{theo}}(A_j)=A_j(2-\pi/2)\exp[-(1-\pi/4)A_j^2]$.
The gradual onset of the Gaussian regime is consistent with the decay of the coherent ballistic component over several scattering mean free paths $\ell_s$, after which $\langle{\bm Z}\rangle\approx{\bm 0}$. This occurs well before the transport mean free path. 

Neither limiting theory is sufficient in the intermediate regime: the paraxial law misses backflow, while the isotropic law misses the finite directed flux.
The eigenvalue evolution in Fig.~\ref{fig:poynting}(e) gives the corresponding physical interpretation. In the present geometry, the initially dominant positive eigenvalue reflects the incident linear polarization. Multiple scattering transfers weight into the second positive eigenvalue, consistent with polarization mixing, while the two negative eigenvalues quantify the emergence of local backflow. Their near degeneracy is consistent with an almost unpolarized backscattered component.
For an initially circularly polarized wave, the polarization memory~\cite{van2015evolution} manifests itself in a slower convergence of the flux eigenvalues, which in that case are related to helicity (see End Matter).

The same eigenvalue law holds for microscopically different disorder.
For a random ensemble of spherical scatterers with refractive-index contrast $\Delta n=0.1$, radius $R=2\lambda$, and volume fraction $\phi=23.56\%$, we find similar agreement between Eq.~(\ref{eq:PSz}) and numerical histograms, although the weaker disorder produces a more skewed distribution and smaller negative eigenvalues.
This comparison, presented in End Matter, demonstrates that Eq.~(\ref{eq:PSz}) accurately describes the statistics of Gaussian vectorial wave fields irrespective of whether the underlying disorder consists of continuous refractive-index fluctuations or discrete scattering particles.
The microscopic realization of the disorder therefore enters only through the covariance matrix.

Remarkably, the derivation above does not rely on any property unique to the Poynting vector.
The essential ingredient is that the local energy flux is a quadratic function of the electromagnetic field amplitudes.
Consequently, the same statistical framework applies to any physical observable that can be expressed as a local quadratic flux of the field.
A relevant example is the recently introduced Fisher-information flux~\cite{Hpfl2024,weimar2026controllingflowinformationoptical}, which quantifies the spatial transport of information about a parameter $\theta$ of a scattering system,
\begin{equation}
{\bm S}^{\rm FI}
=
\frac{2}{\hbar\omega}
\operatorname{Re}
\left(
\partial_\theta{\bm E}
\times
\partial_\theta{\bm H}^{*}
\right).
\label{eq:FI_flux}
\end{equation}
Consider, for example, a coherent laser beam illuminating an object embedded inside a disordered medium [cf.~Fig.~\ref{fig:concept}(c)], where $\theta$ denotes the position or orientation of the object~\cite{Bouchet2021}.
For coherent illumination, the scattering process imprints information about $\theta$ onto the outgoing electromagnetic field, which transports this information from the object of interest through the disordered medium to the detector. The collected Fisher information determines, through the Cram\'er--Rao bound, the highest achievable precision with which the parameter $\theta$ can be estimated~\cite{Bouchet2021,Hpfl2024,weimar2026controllingflowinformationoptical}.
The expression in Eq.~(\ref{eq:FI_flux}) has the same mathematical structure as the Poynting vector, but it is constructed from the sensitivity fields $\partial_\theta{\bm E}$ and $\partial_\theta{\bm H}$ rather than from the electromagnetic fields themselves.
It therefore describes the transport of information about the parameter $\theta$ instead of electromagnetic energy.
The parameter derivative quantifies how sensitively the scattered field responds to infinitesimal changes of the object and serves only to define the Fisher information; experimentally, the corresponding information can be extracted from measurements of the scattered field using an appropriate estimator without physically perturbing the object. When the sensitivity fields, which are sourced at the parameter-dependent region and multiply scattered by the same disorder, have themselves entered a zero-mean circular Gaussian random-field regime, the longitudinal Fisher-information flux obeys the same eigenvalue law,
\begin{equation}
{\bm C}\rightarrow
{\bm C}_\theta=
\frac{4}{\hbar\omega}
\left\langle
\partial_\theta{\bm Z}
\partial_\theta{\bm Z}^{\dagger}
\right\rangle,
\qquad
S_z\rightarrow S_z^{\rm FI}.
\label{eq:replacement}
\end{equation}
The same four-eigenvalue distribution in Eq.~(\ref{eq:PSz}) then applies to the eigenvalues of ${\bm C}_\theta{\bm A}$, where ${\bm C}_\theta$ is the covariance matrix of the rescaled sensitivity vector $2\partial_\theta{\bm Z}/\sqrt{\hbar\omega}$.
Unlike the Poynting flux in a passive homogeneous medium, however, the Fisher-information flux generally possesses sources and sinks wherever the refractive-index distribution depends on the parameter $\theta$.
\begin{figure}[t]
\includegraphics[width=0.48\textwidth]{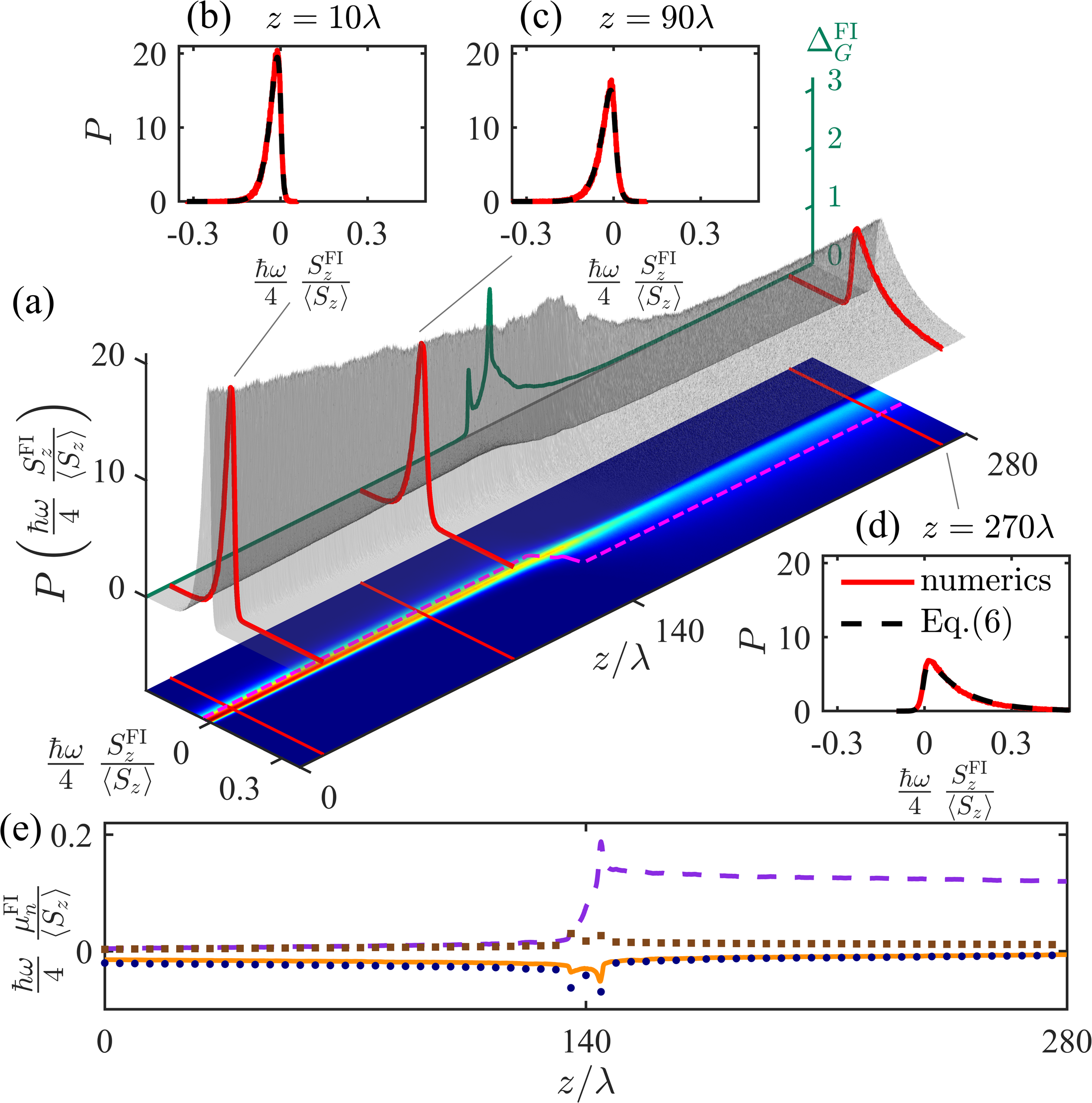}
\caption{
Universal statistics of Fisher-information flux.
(a) A spherical scatterer embedded in a disordered medium displaced by the parameter $\theta=x_{\rm S}$, generating a sensitivity field and an associated Fisher-information flux $S_z^{\rm FI}$ [cf.~Fig.~\ref{fig:concept}(c,d)].
(b--d) Probability distributions of $S_z^{\rm FI}$ at representative planes compared with the eigenvalue law obtained from ${\bm C}_\theta{\bm A}$. We used the average Poynting flux to normalize the Fisher-information flux.
(e) Evolution of the Fisher-information flux eigenvalues.
The dominant tail reverses sign when the observation plane crosses the parameter-dependent scatterer, reflecting the source-sink structure of information flow.
\label{fig:FI}}
\end{figure}

Figure~\ref{fig:FI} illustrates this point for a sphere of refractive index $n=1.8$ and radius $R=5\lambda$ embedded inside a continuous disordered medium with $\sigma=2\lambda$, $\kappa/\sigma^{3/2}=0.1$, $\ell_s=8.2\lambda$, and $\ell_t=2922\lambda$. The disordered slab again occupies $0\leq z\leq L=280\lambda$, and the sphere is placed within this interval. The parameter of interest is the lateral position $\theta=x_{\rm S}$ of the sphere. Since the parameter dependence is localized at the boundary of the sphere~\cite{Hpfl2024,weimar2026controllingflowinformationoptical}, sources and sinks of Fisher-information flux are likewise confined to its surface. Accordingly, the continuity equation for Fisher information \cite{Hpfl2024} implies that the transversely averaged longitudinal Fisher-information flux is conserved everywhere except where the observation plane crosses the sphere, where the mean flux changes due to the localized source term. If the sphere were placed outside the medium (cf.~End Matter), the average information flux would be preserved within the medium. The numerical distributions of $S_z^{\rm FI}$ at representative observation planes are accurately reproduced by the four-eigenvalue distribution of Eq.~(\ref{eq:PSz}) using the eigenvalues of ${\bm C}_\theta{\bm A}$ [Figs.~\ref{fig:FI}(b--d)].
This demonstrates that the statistical theory derived for the Poynting flux carries over unchanged to Fisher-information flux, showing that the local statistics of both energy and information transport are governed solely by the covariance structure of the underlying Gaussian field.

Beyond confirming the universal statistics, the sign of the dominant tail has a direct physical interpretation:
before the observation plane reaches the sphere, information about its displacement can only arrive by propagating backward from the object, giving rise to predominantly negative Fisher-information flux.
After crossing the sphere, the information generated at the object is transported predominantly in the forward direction, and the dominant tail changes sign.
Evidently, if the sphere were outside the medium no such change in direction of the dominant tail would take place (cf.~End Matter).
The Fisher-information flux eigenvalues therefore provide a statistical fingerprint of the directionality of information transport through a complex disordered medium.

Our theory applies whenever the electromagnetic field has entered the Gaussian random-field regime generated by the coherent superposition of many statistically independent scattering paths, as typically encountered in diffusive multiple scattering. This regime does not require statistical isotropy; a finite average Poynting flux is fully compatible with the theory. Conversely, systematic deviations from Eq.~(\ref{eq:PSz}) directly signal the breakdown of zero-mean circular Gaussian field statistics. Such deviations are expected near isolated resonances, in random lasing, or in strongly localized wave regimes, where higher-order field correlations become important. Equation~(\ref{eq:PSz}) therefore provides a natural Gaussian reference against which non-Gaussian energy and information transport can be identified experimentally. Since the flux eigenvalues derive from the local covariance matrix, they should moreover be accessible to transport theory~\cite{Dorokhov1982JETPL,MELLO1988,Beenakker1997RMT,Rotter2017,g3kd-sg4x,djhy-16mh}, making the prediction of $\mu_n(z)$ from radiative-transfer or diffusion models a natural next step. Experimentally, the joint access to transverse electric and magnetic field components required to sample $P(S_z)$ is within reach of vectorial near-field probes and of microwave experiments, where both fields can be measured directly \cite{Hpfl2024,PhysRevLett.134.183802}.

In summary, we have established a universal statistical theory of local energy and information flow in random electromagnetic fields.
Its central result is that the longitudinal Poynting flux in a general nonparaxial Gaussian random field is neither governed by conventional intensity statistics nor by the isotropic random-wave limit.
Instead, its full probability distribution is completely determined by four eigenvalues of the electromagnetic covariance matrix.
These flux eigenvalues provide a compact physical fingerprint of forward transport, optical backflow, and polarization mixing.
The same framework naturally extends to the Fisher-information flux, thereby establishing universal statistics for both energy and information transport in random electromagnetic fields.
\begin{acknowledgments}
We thank J. H{\"u}pfl for helpful feedback.
We acknowledge support from the Department of Precision and Microsystems Engineering and from SURF (www.surf.nl) for access to the National Supercomputer Snellius.
\end{acknowledgments}
\bibliography{reference}

\appendix
\section{End matter}
\textit{Paraxial probability distribution.---}
In the paraxial situation, $H_y$ and $H_x$ are linear combinations of $E_x$ and $E_y$, respectively, and the full field is completely determined by at most two independent complex degrees of freedom. Thus, two eigenvalues $\mu_n$ must be zero. The other two are given by $\mu_n=\left(\frac{1}{p}\langle S_z\rangle, (1-\frac{1}{p})\langle S_z\rangle\right)$ with $p=1$ for the scalar case and $p=2$ the transverse unpolarized case as expected from the trace of $\bm{CA}$.
This yields the well-known expression for the probability $P_{\rm parax}(S_z)$~\cite{Goodman2015} via:
\begin{equation}
    P_\text{parax} =\int\frac{\mathrm{e}^{-\mathrm{i}kS_z}}{\left(1-\frac{\mathrm{i}k\langle S_z\rangle}{p}\right)^p}\frac{dk}{2\pi}= \frac{p^2S_z^{p-1}\mathrm{e}^{-\frac{pS_z}{\langle S_z\rangle}}}{\langle S_z\rangle^p} \Theta(S_z).
    \label{eq:parax}
\end{equation}
Here, $\Theta$ denotes the Heaviside function.

\begin{figure}[tbp]
\includegraphics[width=0.48\textwidth]{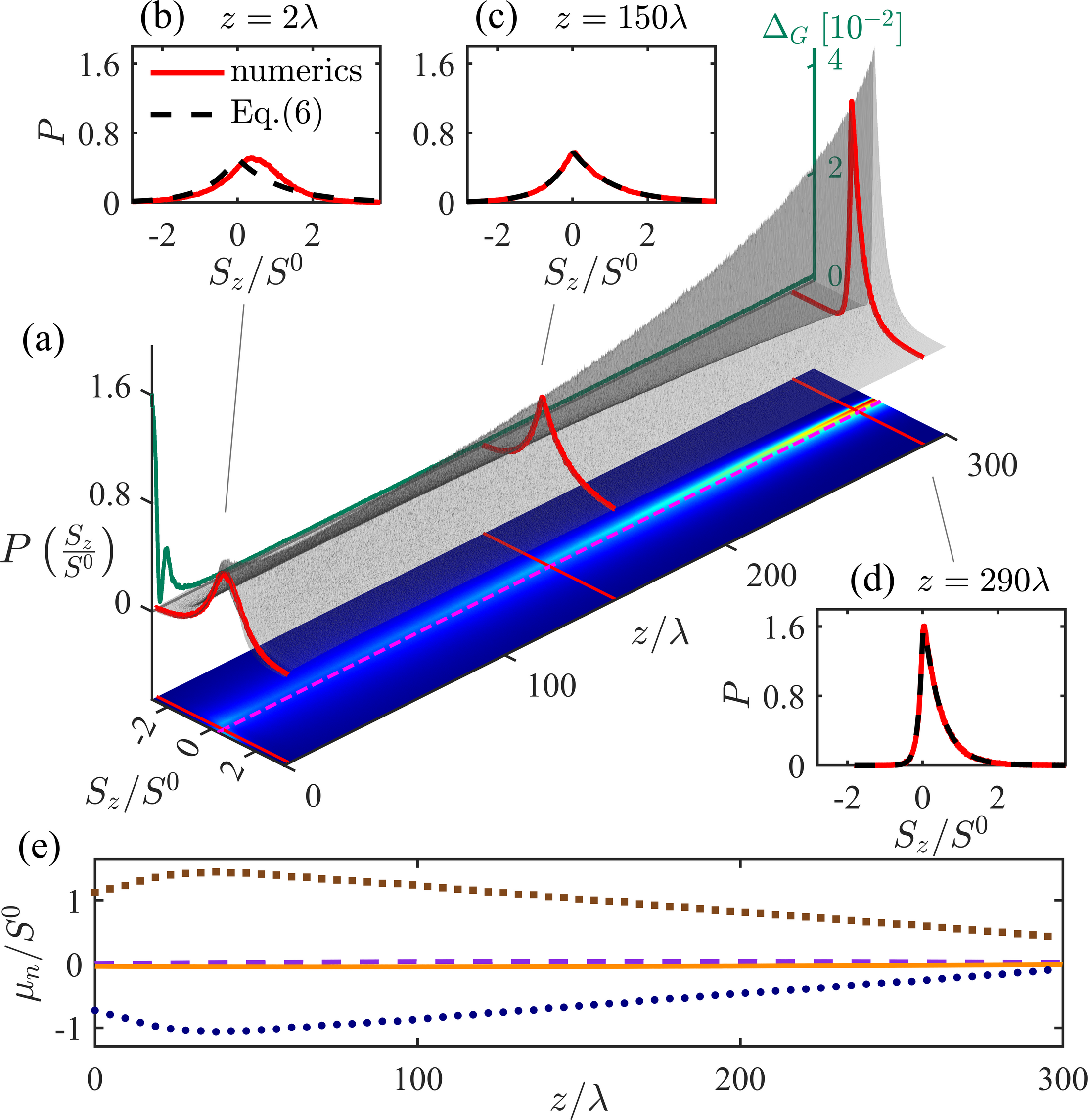}
\caption{
Poynting-flux statistics for circularly polarized incidence in the same medium as Fig.~\ref{fig:poynting}.
The eigenvalue evolution shows slower depolarization than for linearly polarized input.
\label{fig:S1}}
\end{figure}

\textit{Poynting-flux statistics for a circularly polarized incident plane wave.---}
We consider a left-handed circularly polarized incident plane wave in the same continuous disordered medium as in Fig.~\ref{fig:poynting}.
The flux statistics are again described by Eq.~(\ref{eq:PSz}), but the eigenvalue dynamics differs from the linearly polarized case.
Circular polarization depolarizes more slowly, consistent with the known stronger polarization memory of circularly polarized light in scattering environments~\cite{van2015evolution}.
As a result, the convergence of the positive eigenvalues and of the negative eigenvalues is slower than for linear incidence; see Fig.~\ref{fig:S1}.

\begin{figure}[tb]
\includegraphics[width=0.48\textwidth]{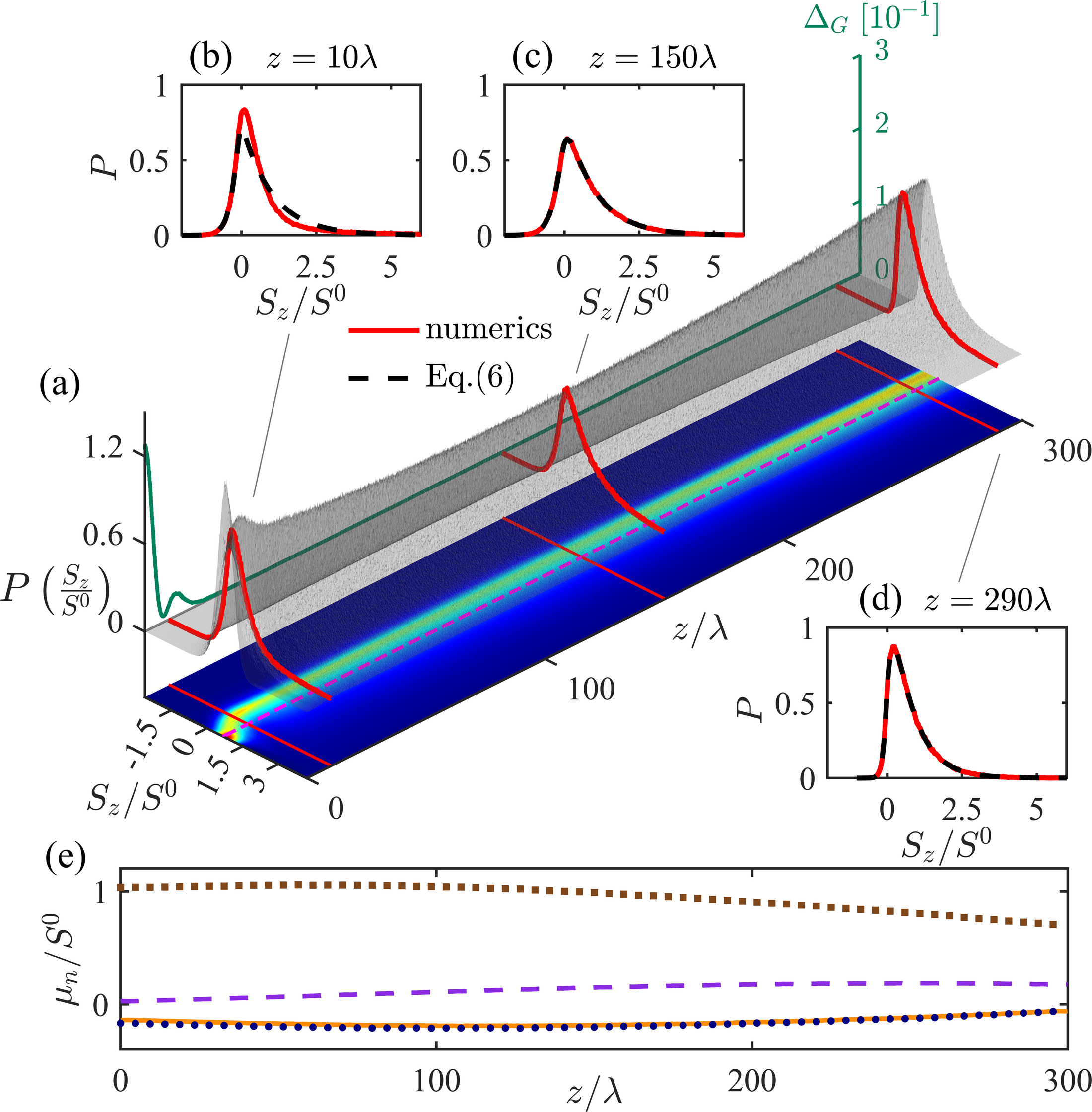}
\caption{
Poynting-flux statistics in a disordered medium composed of discrete scatterers.
The same eigenvalue law describes the probability distribution, confirming that the result is independent of microscopic disorder details once the field is Gaussian.
\label{fig:S_discrete}}
\end{figure}

\textit{Discrete-scatterer medium.---}
To confirm that the eigenvalue law is not tied to a particular disorder model, we also consider a medium composed of discrete spherical scatterers, shown in Fig.~\ref{fig:S_discrete}.
The refractive-index fluctuation is
\begin{equation}
\eta({\bm r})=
\Delta n\sum_{j=1}^{N}
\Theta(R-|{\bm r}-{\bm r}_j|),
\end{equation}
with $\Delta n=0.1$, $R=2\lambda$, volume fraction $\phi=23.56\%$, and background index $n_0=1.5$.
The corresponding scattering and transport mean free paths are $\ell_s=5.8\lambda$ and $\ell_t=373\lambda$.
The distribution remains more skewed than in the stronger continuous-disorder example because backscattering is weaker, but the numerical data are again accurately described by Eq.~(\ref{eq:PSz}).

\begin{figure}[hb]
\includegraphics[width=0.45\textwidth]{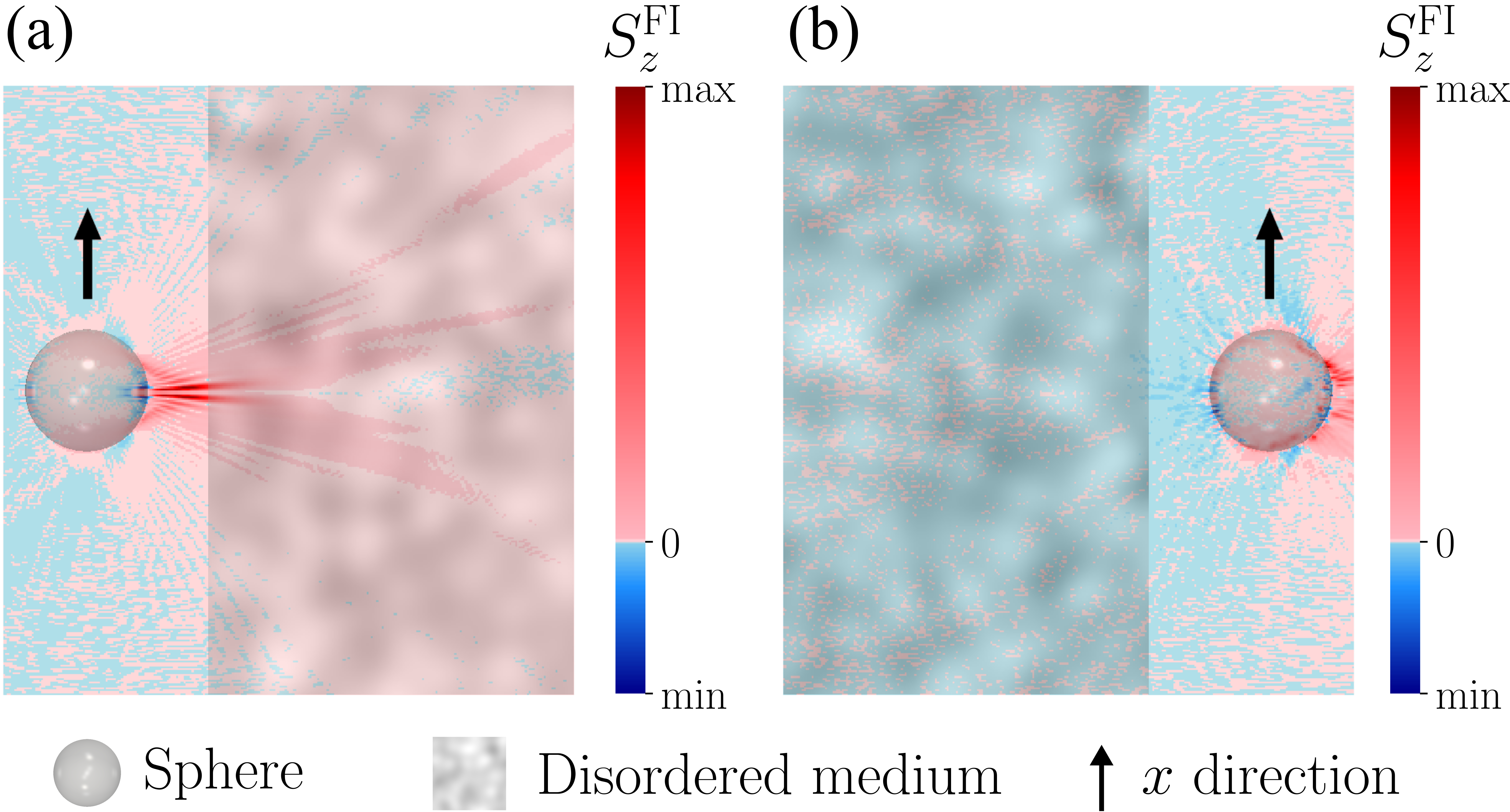}
\caption{
Excerpt of Fisher-information flux distribution for a  sphere placed (a) in front of or (b) behind the disordered medium. The parameter of interest is the lateral position $\theta=x_{\rm S}$ of the sphere.
\label{fig:S4}}
\end{figure}

\begin{figure}[h]
\includegraphics[width=0.48\textwidth]{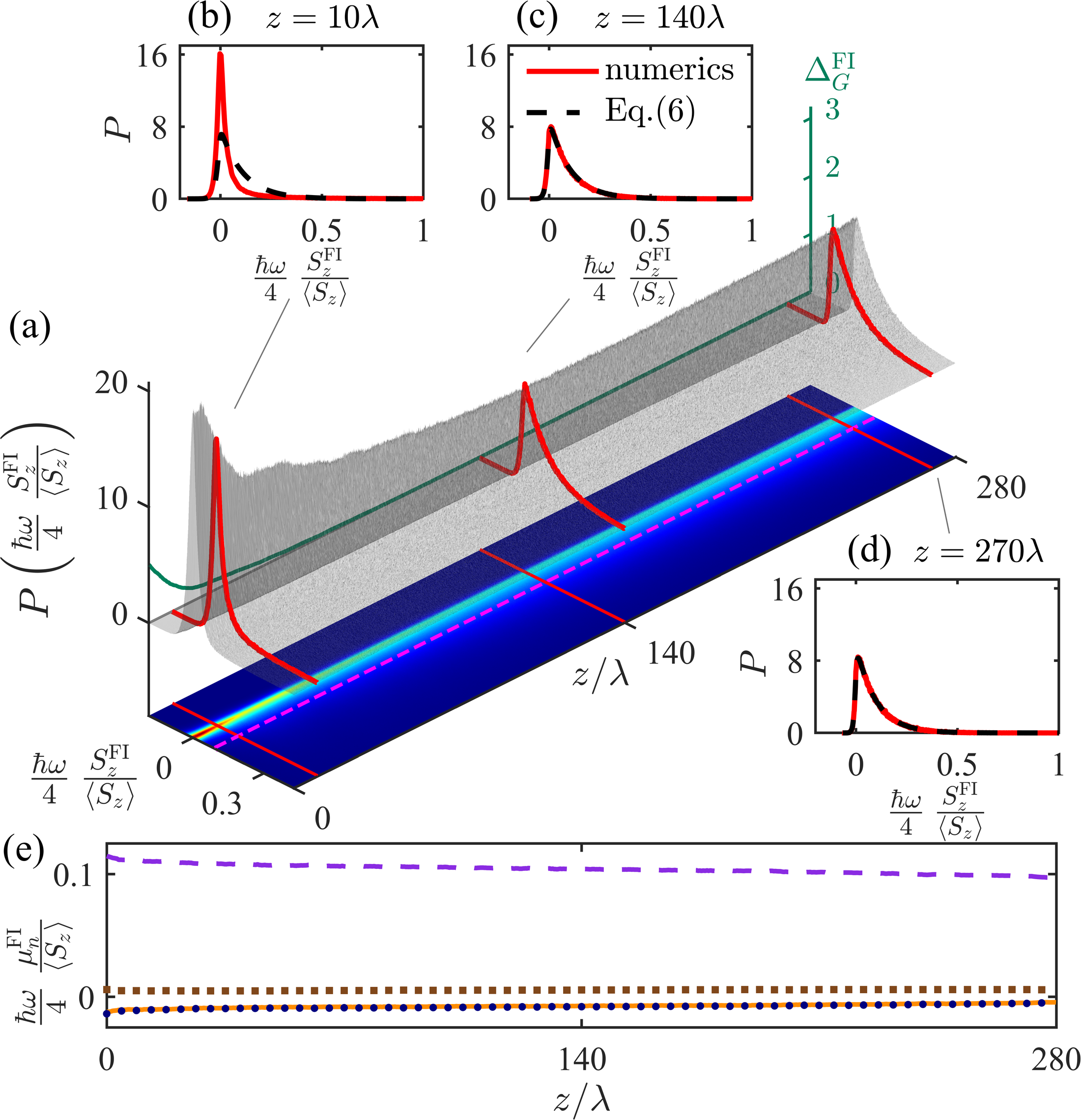}
\caption{
Fisher-information flux for a sphere placed in front of the disordered medium.
The information flux is predominantly forward after the source.
\label{fig:S5}}

\end{figure}

\begin{figure}[htbp]
\includegraphics[width=0.48\textwidth]{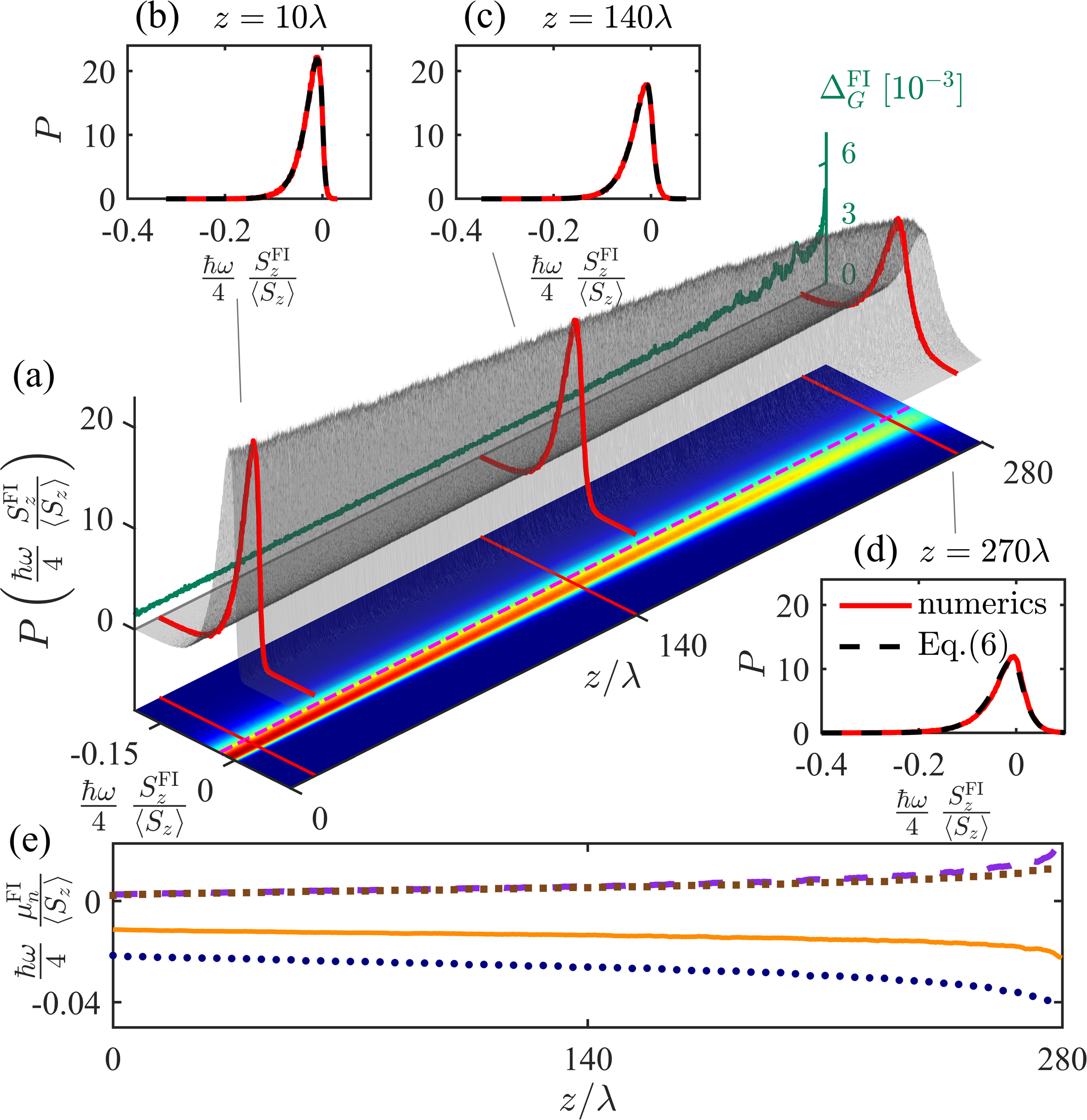}
\caption{
Fisher-information flux for a sphere placed behind the disordered medium.
In front of the sphere, information about the displacement propagates predominantly backward.
\label{fig:S6}}
\end{figure}

\textit{Sphere placed before or after the disordered medium.---}
We also consider the same spherical scatterer placed before (cf.~Fig.~\ref{fig:S4}(a) and Fig.~\ref{fig:S5}) or after (cf.~Fig.~\ref{fig:S4}(b) and Fig.~\ref{fig:S6}) the disordered medium rather than inside it.
When the parameter-dependent object lies outside the disordered medium, the Fisher-information flux is conserved through the disordered region itself.
The sign of the dominant tail depends on whether the observation plane lies before or after the parameter source.

\clearpage
\end{document}